\documentclass[11pt,twoside]{article}
\usepackage{asp2010}

\resetcounters

\begin{document}

\allowtitlefootnote

\title{Equation of state of magnetar crusts from Hartree-Fock-Bogoliubov atomic mass models}
\author{N. Chamel$^1$, R. L. Pavlov$^2$, L. M. Mihailov$^3$, Ch. J. Velchev$^2$, Zh. K. Stoyanov$^2$, Y. D. Mutafchieva$^2$, 
M. D. Ivanovich$^2$, A.F. Fantina$^1$,J.M. Pearson$^4$ and S. Goriely$^1$
\affil{$^1$Institute of Astronomy and Astrophysics, Universit\'e Libre de Bruxelles, CP 226, Boulevard du Triomphe, B-1050 Brussels, Belgium\\
$^2$Institute for Nuclear Research and Nuclear Energy, Bulgarian Academy of Sciences, 72 Tsarigradsko Chaussee, 1784 Sofia, Bulgaria\\
$^3$Institute of Solid State Physics, Bulgarian Academy of Sciences, 72 Tsarigradsko Chaussee, 1784 Sofia, Bulgaria\\
$^4$D\'ept. de Physique, Universit\'e de Montr\'eal, Montr\'eal (Qu\'ebec), H3C 3J7 Canada}}


\begin{abstract}
The equation of state (EoS) of the outer crust of a cold non-accreting magnetar has been determined using the model of ~\cite{lai91}. 
For this purpose, we have made use of the latest experimental atomic mass data complemented with a Hartree-Fock-Bogoliubov (HFB) mass model. 
Magnetar crusts are found to be significantly different from the crusts of ordinary neutron stars.  
\end{abstract}

\section{Introduction}
\label{intro}

Whereas most neutron stars are endowed with typical magnetic fields of order $10^{12}$~G, a few of them have 
been found to have much stronger fields. Huge fields could be generated via dynamo effects in hot newly-born 
neutron stars with initial periods of a few milliseconds. Soft-gamma repeaters and anomalous X-ray pulsars are 
expected to be the best candidates of these so called \textit{magnetars}~\citep{woods2006}. In particular, a 
surface magnetic field of about $2.4\times 10^{15}$~G has been inferred in SGR~1806$-$20 and its internal field 
could be as high as $10^{18}$~G~\citep{lai91}. The presence of such strong fields makes the properties of 
magnetar crusts very different from those of ordinary neutron stars. 

\section{Structure and equation of state of magnetar crusts}
\label{eos}

\begin{figure}
\centering
\includegraphics[scale=0.33]{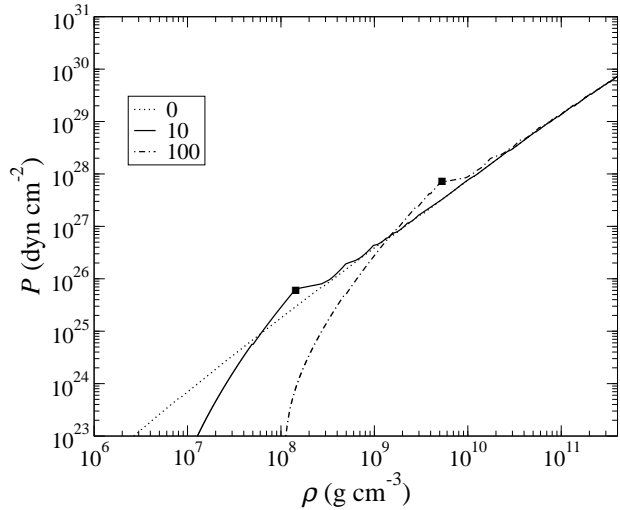}
\caption{Pressure $P$ vs mass density $\rho$ in the outer crust of a cold non-accreting neutron star for different magnetic field strengths 
(in units of $B_c$). The filled squares indicate the points at which the lowest Landau level is fully occupied.}
\label{fig1}
\end{figure}

We have studied the structure of the outer crust of a cold non-accreting magnetar using the model of 
~\cite{lai91}. For this purpose, we have made use of the most recent experimental data from a preliminary unpublished 
version of an updated Atomic Mass Evaluation. For the atomic masses that have not yet been measured, we have employed the 
microscopic model HFB-21 of ~\cite{goriely2010}. In a strong magnetic field, the electron motion perpendicular to the field 
is quantized into Landau levels. As a result, the equilibrium composition of magnetar crusts can significantly differ from that 
of ordinary neutron stars, especially when the magnetic field strength $B$ exceeds 
$B_c\equiv m_e^2 c^3/(e\hbar)\simeq 4.4\times 10^{13}$~G~\citep{rila2011}. For instance, $^{66}$Ni which is found in the outer 
crust of neutron stars for $B=0$~\citep{pearson2011} disappear for $B>67 B_c$. On the other hand, $^{88}$Sr is only found in 
magnetar crusts for $B> 859 B_c$. Moreover, strong magnetic fields prevent neutrons from dripping out of nuclei. As a result, 
the pressure at the neutron drip transition increases from $7.82\times 10^{29}$ dyn~cm$^{-2}$ for $B=0$ to 
$1.05\times 10^{30}$ dyn~cm$^{-2}$ for $B=1000 B_c$. This might have implications for the interpretation of pulsar glitches~\citep{lrr}. 
As shown in Fig.~\ref{fig1}, the strongly quantizing magnetic fields prevailing in magnetar interiors have a large impact on the EoS 
in the regions where only a few Landau levels are filled. With increasing density, the effects of $B$ become less and less 
important as more and more levels are populated and the EoS matches smoothly with that 
obtained by~\cite{pearson2011} for $B=0$. 

\acknowledgments 
This work was supported by FNRS (Belgium), NSERC (Canada), Wallonie-Bruxelles-International (Belgium) and the 
Bulgarian Academy of Sciences. 

\bibliography{chamel_poster}

\end{document}